\documentclass[11pt]{article}
\usepackage[dvips]{epsfig}
\begin{document}

\begin{center}
{\bf A Kalman Filter for Ocean Monitoring\\}

\vspace{6.mm}
Konstantin P.Belyaev\\

\vspace{3.mm}
{\it LNCC, Petropolis, Brazil\\}

\vspace{6.mm}
Detlev M\"{u}ller\\

\vspace{3.mm}
{\it MPIMET, Hamburg, Germany\\}

%\vspace{6.mm}
%August 2002\\
\end{center}

\vspace{10.mm}
{\small
{\bf Abstract:} The feasibility of global ocean state estimation by sequential 
data assimilation is demonstrated. The model component of the assimilator is the
GROB version of the MPIMET ocean circulation model HOPE. Assimilation uses the
Fokker-Planck representation of the Kalman Filter. This approach determines the
temporal evolution of error statistics by integration of the Fokker-Planck
Equation. Phase space advection and diffusion are obtained from histogram
techniques considering the model as a black box. For efficiency, the estimation
procedure utilizes a combination of nudging and Kalman Filtering. The ocean state
is estimated for the El Nino year 1997 by dynamical extrapolation of observed
sea-surface temperatures and TAO/TRITON subsurface temperatures. The model-data
combination yields improved estimates of the ocean's mean state and a realistic
record of El Nino related variability. The assimilator proves as an efficient,
viable and thus practical approach to operational global ocean state estimation.}

\vspace{10.mm}
\noindent
{\bf 1.) Introduction\\}

Ocean state estimation draws its wider societal as well as scientific significance from the 
central role of the oceans in Earth's climate system. At a time when the impact of climate 
variability on societal infrastructures is increasingly felt, the need for comprehensive 
climate monitoring is generally accepted. While mankind is primarily affected by 
meteorological manifestations of climate variability, large-amplitude weather fluctuations 
oftentimes screen the atmospheric climate signal. In practice, atmospheric climate 
observation proves prohibitively intricate. Alternatively, estimates of the state of the 
ocean interior with its enormous capacity to store and distribute water, heat and radiatively 
active trace substances (such as carbon dioxide) provide direct evidence of the climate signal. 
As the dominating climate component, the ocean acts as a Markov Integrator of atmospheric 
noise, provides the memory of the climate system and sets time-scales of climate processes 
by (at least partly) predictable transport mechanisms. Thus, practical climate monitoring 
anchors on operational ocean state estimation. 

Only a decade ago, an observational basis for the assessment of a temporally changing global
ocean circulation was practically nonexistent. In fact, oceanography was plagued by a
notorious undersampling problem and the ocean circulation was widely considered as a
stationary flow. The most comprehensive data set of this era is the World Ocean Atlas (WOA) 
by Levitus and coworkers \cite{1}, a meticulous collection of hydrographic data
from oceanographic archives all over the world. This atlas represents a global ocean density 
field that is best associated with the mean ocean circulation during the second half of the 
twentieth century. Temporal changes such as those at the roots of past, present and future 
climate variability are far beyond the scope of these data. Since then, ocean observation 
underwent an explosive development. Today, space-borne ocean observatories provide global data 
sets of the mesoscale state of the sea surface in near real-time. Oceanographic data paucity 
has been replaced by an almost overwhelming data stream \cite{2}. Nevertheless, 
observational 
monitoring of the ocean interior remains technically difficult and costly for the foreseeable 
future. For mere book keeping as well as for analysis and interpretation, ocean state 
estimation thus relies heavily on the dynamical extrapolation of observations by means of 
numerical models of the global ocean circulation. 

Such models play a central role in the study of the climate system and its ocean component for 
more than three decades. As engineering devices with the capacity for otherwise impossible 
experimentation they provide a laboratory for the simulation, extrapolation and ultimately 
understanding of the observational record. So far, the most convincing demonstration of the 
potential of circulation modeling has been the discovery of the El Nino mechanism and its 
repercussions on remote regions of the globe. After Matsuno's theoretical ground work 
\cite{3} and the provision of observational evidence by Wyrtki \cite{4}, the essential 
dynamical processes were numerically simulated with simple shallow water models \cite{5,6}.
Today, operational El Nino monitoring and forecasting are implemented at a number of 
institutions world wide and the variability of tropical circulations has become one of the 
best understood aspects of the climate system.

On the other hand, dedication of more complex models to the same problem has yet been unable
to advance El Nino simulation and forecasting significantly. Nor has it been possible to gain 
comparable insights for the extratropics. While scientists are well aware of extratropical
circulation variability such as the North Atlantic or Arctic Oscillation or the Antarctic
Circumpolar Wave, numerical analyses have clearly been less yielding in these instances.
Uncertainties of this kind also raise the question of the adequacy of contemporary numerical
circulation models for longer term climate projections.

The major difficulty of global circulation modeling is the lack of a physically consistent
and numerically soluble formulation of the circulation problem. Physically, the global
ocean circulation poses the thermohydrodynamical problem for a viscous fluid on the
rotating geoid. The gravest problem of circulation theory is certainly the absence of an
energetically consistent formulation of the equations of motion for a viscous rotating
fluid. While such formulations are well known for viscous nonrotating fluids and
ideal rotating fluids, the derivation of the energy budget from the equations of motion
of a dissipative rotating fluid remains a matter of scientific debate. As one consequence, 
wave-dissipation in rotating fluids is not very well understood. Moreover, viscosity
plays a significant role for the numerical stability of circulation models. It is clearly
desirable that parameterizations of subscale transports steer momentum, energy and vorticity
along realistic, i.e. energetically and vortically consistent pathways in space-time and
wave vector space. For contemporary circulation models, this matter remains essentially 
unsolved. 

Satellite orbits reveal a complex fine structure of the planet's shape and considerable 
uncertainties particularly about the marine geoid still exist. For most purposes of 
circulation modeling, however, an approximation in terms of a spheroid or even a sphere is
probably sufficient. A source of dissatisfaction with 3-dimensional spheroidal or spherical 
circulation equations are difficulties in analytically obtaining simple stationary solutions 
which reflect characteristic circulation features such as geostrophy and thermal wind 
balance. Moreover, linearizations in these coordinates do generally not admit separation of 
variables. Hence, it has yet been impossible to study analytically the propagation of 
acoustic, gravity and Rossby wave disturbances together with the stability of simple flows 
in a common, 3-dimensional framework. For numerical integration, such problems are 
insignificant. Nevertheless, most contemporary numerical circulation models compromise 
geometric-dynamic integrity in favor of a multi-$\beta$-plane approximation to the geometry 
of the geoid.  This approach codes Laplace operators as sum of second order derivatives and 
ignores first order contributions from nontrivial Christoffel symbols.

A third set of issues of circulation modeling is associated with the nonlinearity of fluid
dynamics. Nonlinear field theories inevitably couple variability on the smallest space- and
time-scales with the largest scales available. For numerical as well as theoretical
purposes, processes on small spatial and fast temporal scales should be eliminated from
circulation equations. In the first place, this applies to acoustics. The widely favored 
sound filter invokes the Boussinesq approximation to inertia and weight of the fluid. While 
this approach has been quite successful in the study of internal gravity waves and 
Rayleigh-Benard instability, its
vorticity-inconsistency becomes a problem in long-term integrations of circulation
modeling. Furthermore, Rayleigh-Benard (or static) instability involves fast convective
motions on small spatial scales. Such convective events are generally not resolved in
global circulation models and their crucial role for the thermohaline circulation is
represented by appropriate parameterizations. Hence, models assume the ocean to be in 
hydrostatic equilibrium and simply neglect internal vertical accelerations relative 
to Earth's gravitational acceleration. However, this straightforward introduction of 
hydrostatics generates a problem. Now, momentum density is a 2-dimensional vector while 
the mass flux vector of the continuity equation remains 3-dimensional. Such a violation 
of the first law of motion will generally be uncritical for stationary flows or 
order-of-magnitude estimates. However, for the long-term integrations of circulation 
modeling, such a formulation is not entirely satisfactory.

The Boussinesq approximation, hydrostatics and the so-called ``traditional approximation'' of
inertial forces define Richardson's Primitive Equations \cite{7}. Currently, these equations 
provide the physical basis for most global circulation models. While the Primitive Equations
account consistently for velocities and thermodynamics of equilibrium circulations, they do
not pose a Newtonian dynamical problem.

The present study utilizes GROB HOPE, a coarse version of the numerical Hamburg Ocean 
Primitive Equations model of MPIMET \cite{8}. This model is based on a 
C-grid discretization of the Primitive Equations for UNESCO seawater \cite{9} and allows 
various convection parameterizations which account for different characteristics of this
process in the open ocean and in the bottom boundary layer down submarine slopes. 
GROB HOPE includes sea ice dynamics with viscous-plastic rheology \cite{10} 
parameterizing cracking, ridging, rafting and deformation of sea ice. The model is forced by 
buoyancy fluxes and wind stresses at the sea surface as well as the freshwater discharges of 
Earth's 50 largest rivers.
 
In long term experiments (integration time: 1000 years) with climatological forcing resolving
the annual cycle, the model assumes an essentially drift-free cyclostationary state after a 
few centuries which reproduces the major water masses and gyre structures of the 
global ocean circulation as well as the sea ice cover and its seasonal variation at high 
latitudes. While this model circulation exhibits the characteristic degree of realism of 
state-of-the-art simulations it also displays a number of typical deficits. The model fails
to maintain the observed Pacific Intermediate Waters. Furthermore, while the poleward Atlantic
heat transport is certainly of the observed order of magnitude, its maximum of 0.8 PW is still
somewhat lower than the 1.1 PW suggested by observations. On the other hand, the mass transport
by the Antarctic Circumpolar Current with 180 Sverdrup in the Drake Passage is higher than the 
observed 140 Sverdrup. The path of the Gulf Stream which is crucial for the European climate
and weather turns out to be quite sensitive to the details of the atmospheric forcing and the
chosen parameterization of subscale transports. For an extensive discussion of the strengths
and weaknesses of the GROB HOPE circulation, see \cite{8}. At this time, the 
versatility of numerical circulation models is sufficiently developed to simulate a wide 
variety of preconceived scenarios. However, beyond this illustrative role, state-of-the-art 
models generally lack the capacity of scientific discrimination between competing predictions
and projections. To a large part, these uncertainties can be resolved by the systematic
combination of models with observations.

Such combination and confrontation of models with extensive and novel observational
data sets has been the outstanding factor in forecast-improvement by numerical weather
prediction over the last thirty years. With maturing numerical ocean circulation models and
a growing understanding of the variability of the ocean circulation this approach has now 
also become attractive to the oceanographic and climate communities. Given the characteristics 
of model data and observations in Earth System Modeling, the integration of information from 
different
sources poses a considerable data engineering problem. Frequently, observations do not
refer to prognostic model variables while other parameters such as vertical velocities are
practically unobservable. Moreover, observations are typically distributed highly irregular
in space-time. And data sets from both sources, model and observation, are large. The 
mathematics of the optimizing synthesis of large data sets are the objective of estimation 
theory \cite{11}. To avoid the mutual enhancement of model- and data-error, estimation
theory has been (and still is) developing a number of what are called data assimilation
algorithms. Generally, these algorithms fall into two classes: variational and sequential
techniques \cite{12,13,14}. The equivalence of both methods is readily demonstrated in 
simple cases. 

Variational assimilation, namely the Adjoint Method, is based on an application of inverse 
modeling techniques to the estimation problem. Variation of control parameters minimizes a 
cost function formed by the model-data misfit. This approach lends itself particularly to 
the estimation of equilibrium states and processes of finite duration. Computation of the 
cost gradient with respect to the controls calls for what is often referred to as the
temporally backward integration of the adjoint model. For complex models, 
coding of the model adjoint is a substantial task, well comparable to coding the model 
itself. The practical relevance of adjoint assimilation in Earth System Modeling therefore 
arose only after the advent of the theory of automatic differentiation \cite{15} and 
the subsequent development of automatic adjoint code compilers \cite{16}. With global
circulation models, the Adjoint Method finds presently wide application in sensitivity
studies.   

Sequential methods such as the Kalman Filter are more specifically tailored to the needs of 
monitoring and prediction \cite{17}. These updating schemes emerge from the application of 
the theory 
of stochastic processes to estimation and yield an estimate of minimum variance. On update, 
the relative weight of model and data is determined by the Kalman Gain which is computed 
from data and model error dynamics. To this end, the model error is considered as a
stochastic process. The dynamics of such processes can be equivalently formulated in the
Langevin (or Heisenberg) representation and in the Fokker-Planck (or Schr\"{o}dinger) 
representation \cite{18}. The Langevin picture addresses the space-time behavior 
of the process in terms of its moments. In practice, this refers generally to the covariance 
only. Formally, the temporal development of the covariance is uniquely determined by the model 
dynamics. However, the practical derivation of the covariance dynamics for a complex model 
such as a global ocean circulation model readily becomes everything but straightforward. 
This applies particularly to nonlinear models, the issue of boundary and initial conditions 
for the covariance, stability questions and the problem of temporally backward assimilation. 
Nevertheless, at this time the literature on Kalman Filter assimilation in Earth System 
Modeling and other branches of  engineering is almost exclusively dominated by the Langevin 
approach \cite{14}.

Alternatively, a stochastic process may be considered in phase-space in terms of its 
probability density. Provided the process is Markovian and jumps remain small in an 
appropriate sense \cite{18} the dynamics of this probability density are governed 
by the Fokker-Planck Equation. The advection- and diffusion-coefficients of this linear
parabolic differential equation are determined by model dynamics and observational error
statistics. In general, these coefficients are also difficult to obtain from a complex
model. However, for sufficiently short update intervals, phase-space advection and
diffusion can be determined phenomenologically from the model output by histogram 
techniques. In this framework, the assimilation method provides practical answers to the
issues of phase-space reduction, model nonlinearity, initial and boundary conditions for
higher moment dynamics as well as stability. Moreover, the existence of the Backward 
Fokker-Planck Equation \cite{18} will permit the generalization of sequential 
assimilation to include the temporally backward extrapolation of data information. The 
mathematical aspects of the Fokker-Planck representation of sequential Kalman Filter 
assimilation have been developed in detail by \cite{19}.

With the typical volume of model output and observational record in Earth System Modeling, 
computational demands for assimilation with least-square optimality are always quite high.
For a reduction of the computational burden, the present estimation utilizes a combination
of Kalman Filter assimilation and simple ``nudging''. While subsurface temperatures from
the TAO/TRITON  array will be assimilated sequentially, observations of global sea-surface
temperatures are essentially inserted into the model at daily intervals. The feasibility
of this simplistic technique is by no means trivial. Older model generations were
generally unable to ``digest'' essentially unprocessed data and model-data inconsistencies
would readily emerge in various regions of space-time and phase space. It will here be
shown that the quality of contemporary models and data sets is sufficiently high for
nudging to be beneficial for the ocean state estimate. 

\vspace{10.mm}
\noindent
{\bf 2.) Fokker-Planck Picture of the Kalman Filter.\\}

Societal and scientific needs in Earth System Monitoring are presently best met by an 
operational, steady combination and confrontation of substantial, yet incomplete observations 
with complex, but nevertheless approximate numerical models. This combination aims to fill 
data gaps by dynamical extrapolation of observations on the basis of physical laws 
incorporated in the model and simultaneously constrain model uncertainties by operating the 
model in close vicinity of observations. One source of information is used to compensate 
the deficits of the other and thus arrive at a comprehensive state estimate including
an assessment of model- and observation-quality. The assimilation algorithms of estimation
theory are information integration techniques which prevent the mutual enhancement of errors
from different sources and maximize the benefit from imperfect models and data. As an 
engineering tool, data assimilation accepts or rejects a hypothesis (the model) while
a constructive evaluation of structural model deficits remains beyond its scope.

\begin{figure}[!h]
\centerline{\hbox{
\psfig{figure=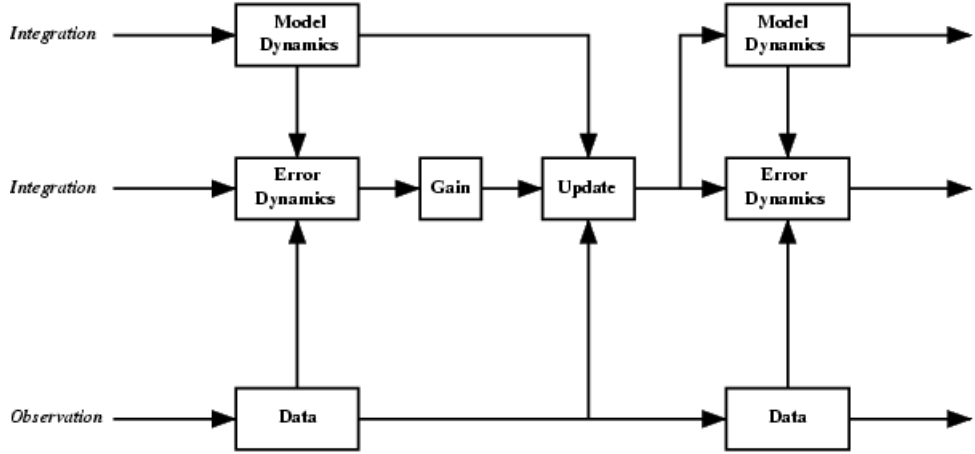,height=6.cm,angle=0,clip=}
}}
\caption{Flow Chart for Sequential Data Assimilation}
\end{figure}

The key idea of sequential assimilation is to integrate the model until an observation becomes 
available. At this time, model integration is halted and the state of the system is updated by 
an appropriate combination of model prediction and observation. Subsequently, this update 
provides the initial value for the continued model integration (fig.1). Hence, the main task 
in sequential data assimilation is the determination of the temporal development of the 
relative weight of model prediction and observation.

The state of the system under consideration (here: the global ocean) is given by a finite- 
(generally: very high) dimensional vector $Y(t,x)$ where $t$ denotes time and 
$x=(x_{(1)},x_{(2)},x_{(3)})$ the 3-dimensional spatial coordinate. In numerical ocean models, 
time and space coordinates will generally be members of a discrete, 4-dimensional grid. The 
time-development of the state of the system is governed by the model dynamics
\begin{equation}
\frac{d}{dt}Y_{m}(t)=\Lambda(Y_{m},t)
\end{equation}
where the index m refers to the model and $\Lambda$ is a generally nonlinear operator of $Y_{m}$
with 
\[\Lambda(Y_{m}=0,t)=0\]
and some explicit time-dependence representing external forcing. For simplicity, spatial 
coordinates have been suppressed in (1). It is now assumed that the model is sufficiently 
``good'' so that the time-development of the true, i.e. observed system is given by the equations
of motion
\begin{equation}
\frac{d}{dt}Y_{o}(t)=\Lambda(Y_{o},t)+W_{0}
\end{equation} 
where the index $o$ refers to observation, while $\Lambda$ is the same operator as in (1) and 
$W_{0}$ an additive noise of known probability distribution. The possibly nonstationary noise is 
assumed to have zero average, finite spatial and no temporal correlation, i.e. it is assumed 
to be temporally white 
\[<W_{0}(t,x)>=0, \hspace{10.mm} <W_{0}(t,x)W_{0}(t+\tau,x+r)>=Q(t,r)\delta(\tau).\] 
Physically, this noise accounts for model deficits. Equation (2) is the Langevin representation 
of a (nonlinear) stochastic differential equation.

The model error
\[y(t,x)=Y_{o}(t,x)-Y_{m}(t,x)\]
appears under a variety of names in different applications of assimilation theory. Variational
techniques generally term this quantity ``misfit'' while numerical weather prediction refers to
the model error as ``innovation''. By definition, the model error satisfies the  Langevin 
equation
\[\frac{d}{dt}y(t)=\Lambda(y+Y_{m},t)-\Lambda(Y_{m},t)+W_{0}.\]
With the help of the mean value theorem, the dynamical operator is rewritten as
\[\Lambda(y+Y_{m},t)-\Lambda(Y_{m},t)=y\Lambda'(Y_{m},t)+R\]
where the residual R comprises terms higher than second order. Writing now
\[\Lambda(y+Y_{m},t)-\Lambda(Y_{m},t)=\Lambda(y,t)-\Lambda(y,t)+y\Lambda'(Y_{m},t)+R\]
and using again the mean value theorem on $\Lambda(y,t)$ together with $\Lambda(0,t)=0$, one 
arrives at 
\begin{equation}
\frac{d}{dt}y(t)=\Lambda(y,t)+W
\end{equation}
where the new noise $W$ comprises $W_{0}$ and terms of second and higher order. It is readily 
seen that
\[<W(t,x)>=0.\]
The error dynamics (3) account for nonlinearities and provide the basis for the histogram
technique to be invoked below.

For simplicity, data are here assumed to represent prognostic model variables directly. 
Furthermore, measurements are made at specific observation points $X=(X_{(1)},X_{(2)},X_{(3)})$ 
which are generally much less in number than model grid points $x$ and do not coincide with 
these. Interpolation to model points will generally be straightforward and observation points 
are here assumed to coincide with model points. In the context of monitoring and prediction, 
observations typically become available at specific times which are termed ``analysis times'' 
in numerical weather prediction. At these times, sequential data assimilation halts the model 
integration and updates the system's state vector with observational information. This update 
is given in terms of a convolution-type integral 
\begin{equation}
Y(t,x)=Y_{m}(t,x)+\int_{0}^{t} d\tau \int dX\, G(\tau,x,X)\,y(\tau,X)
\end{equation}
where the kernel $G(\tau,x,X)$ is called the gain. For the update (4), the gain accounts for 
all observations prior to analysis time and hence only represents the temporally forward 
propagation of data information. In this sense, the kernel
in (4) represents the ``retarded gain''. It is equally possible to ask: at which state did
the system initially start, if it is at the observed state on analysis time. The account of
the temporally backward propagation of data information requires the inclusion of the 
``advanced gain''. Sequential assimilation admits the dynamical extrapolation of data 
information into both, past and future. The corresponding algorithm, namely a meaningful 
representation of the advanced gain becomes particularly transparent in the phase space 
representation. Its formal details will be discussed elsewhere.

There is a number of expressions for the gain-kernel in (4) which all lead to a different variance
for the state estimate. The estimate has minimum variance if the gain is given by the Wiener-Hopf 
equation
\begin{equation}
K(t,x,X)=\int_{0}^{t} d\tau \int dX_{0}\,G(\tau,x,X_{0})\,K(t-\tau,X,X_{0})
\end{equation}
where
\[K(t,x_{1},x_{2})=<y(t,x_{1})y(t,x_{2})>-<y(t,x_{1})><y(t,x_{2})>.\]
is the error covariance with brackets indicating the ensemble mean and unbraced indices denoting 
different points, i.e. different 3-tuples (and not different vector components). The gain defined
by equation (5) is called the Kalman Gain and the sequential data assimilation algorithm given by
(4) and (5) is called the Kalman Filter \cite{20}. It follows from (5) that the central 
problem of Kalman Filter assimilation is the determination of the temporal development of the 
error covariance $K(t,x_{1},x_{2})$.

Formally, the covariance matrix and its temporal behavior are uniquely determined by the model
error dynamics (3). In practice, however, exploitation of this equation encounters a number of
serious problems. Covariance dynamics on the basis of (3) are determined at every model grid 
point. If model integration requires $N$ operations per time-step, integration of the covariance 
dynamics requires additional $N^{2}$ operations. An order of magnitude estimate of the 
number of operations for a contemporary ocean circulation model such as GROB HOPE is 
$N\approx 10^{5}$. Even taking into account the rate of increase in available computing 
capacities, integration of the complete covariance dynamics in the Langevin picture becomes 
prohibitive for global circulation models. Furthermore, these models typically involve strong 
nonlinearities. For mean quantities such as the the error covariance, this nonlinearity leads to 
a hierarchy problem since the mean of a product does not equal the product of the means. 
The difficulties and ambiguities of practical closures of such hierarchies are well known from 
turbulence theory \cite{21}. Additional complications 
arise from the lack of initial and boundary conditions for the dynamics of higher moments and
for the determination of stability and uniqueness conditions of the assimilation procedure.
These problems represent some of the reasons for the prevalence of the Adjoint Method in Earth
System estimation.

An alternative approach to the Kalman Filter utilizes the phase space or Fokker-Planck
representation of stochastic processes. In this framework, a detailed assimilation algorithm 
for application in Earth System estimation has been developed by Belyaev and coworkers 
\cite{19,22}.  The starting point of this formulation is the joint probability distribution
\[p=p(t, \eta_{1}, \eta_{2}; X_{1}, X_{2})\]
for the two-component projection $\eta$ of the model error $y$
\[\eta=(\eta_{1}, \eta_{2})=(y(t,X_{1}), y(t,X_{2}))\]
to have the value $\eta_{1}$ at observation point $X_{1}$ and the value $\eta_{2}$ at 
observation point $X_{2}$ at time t. In terms of this probability the error covariance (5) takes
the form
\begin{equation}
K(t,X_{1},X_{2})=\int d^{2}\eta\,\eta_{1}\,\eta_{2}\,p-
\int d^{2}\eta\,\eta_{1}\,p\int d^{2}\eta\,\eta_{2}\,p.
\end{equation}
Determination of the error covariance thus requires the calculation of joint probability
distributions p and this operation sets the magnitude of necessary computing capacities. Notice 
that (6) defines the error covariance only for pairs of observation points. In Earth System 
estimation, the number $M$ of these points is typically orders of 
magnitude smaller than the number $N$ of model grid points. Using the symmetry of the joint
probability distribution, the covariance matrix for $M$ observations has $M(M+1)/2$ independent
members. Given 1000 data points (which still is a fairly small data set in Earth System 
observation), the error covariance calls for the computation of half a million probability
functions. While this number is clearly much smaller than $N^{2}\approx 10^{10}$, it still poses
a considerable computational task.

In the phase space picture, the error $\eta$ is viewed as a Markov process. The temporal 
development of the conditional probability distribution of such a process is governed by the
Master Equation \cite{18}. For Markov processes with small jumps this equation is
well approximated by the Fokker-Planck Equation \cite{18,23,24}. In the present case this 
equation takes the form
\begin{equation}
\partial_{t}p=-\partial^{n}\Lambda_{n}p+\frac{1}{2}\partial^{n}(q_{mn}\,\partial^{m}p)
\end{equation}
where the usual summation convention for indices $m,n,\ldots =1,2$ is implied. The vector
\[J_{n}=\Lambda_{n}p-\frac{1}{2}q_{mn}\,\partial^{m}p\]
denotes the advective-diffusive probability flux in phase space with advection velocity 
$\Lambda_{n}$ and diffusion $j_{n}=-\frac{1}{2}q_{mn}\partial^{m}p$. In this sense, the 
Fokker-Planck Equation (7) expresses the conservation of probability in phase space. Temporally, 
this linear parabolic equation governs the development of the conditional probability 
$p(t_{2},\eta_{2}\,|t_{1},\eta_{1})$ of the error $\eta$ to have the value $y_{2}$ at time 
$t_{2}$ and location $X_{2}$ given its value $y_{1}$ at time $t_{1} < t_{2}$ at location 
$X_{1}$. The required joint probability follows from the solution of the Fokker-Planck Equation
by Bayes' Rule
\[p(t,\eta_{2},\eta_{1})=p(t,\eta_{2}\,|\,t,\eta_{1})p(t,\eta_{1}).\]
It is noted that there is also a Backward Master Equation \cite{18,24}. The small jump 
approximation to this equation leads to the Adjoint Fokker-Planck Equation which governs the 
temporally reversed development of the error distribution. Hence, the Adjoint 
Fokker-Planck Equation provides the basis for the determination of the Advanced Kalman Gain.

For the solution of the Fokker-Planck Equation (7) the phase space advection 
$\Lambda_{n}(t,\eta)$ and the diffusion tensor $q_{mn}(t,\eta)$ have to be known. These 
parameters are determined by error dynamics and data stochastics according to
\[\Lambda_{n}(t,\eta)=<\Lambda_{n}(t,\eta)\,|\,\eta=y>=
\tau^{-1}\int d^{2}\eta'\,(\eta_{n}-\eta_{n}')\,p(t,\eta\,|\,t',\eta')\]
with $\tau=t-t'$ for $t'<t$ for the advection while one has for the diffusion tensor
\[q_{mn}(t,\eta)-Q_{mn}(t,\eta)=
<\Lambda_{m}(t,\eta)\Lambda_{n}(t,\eta)\,|\,\eta=y>=
\tau^{-1}\int d^{2}\eta'\,(\eta_{m}-\eta_{m}')(\eta_{n}-\eta_{n}')\,p(t,\eta\,|\,t',\eta')\]
where $Q_{mn}$ denotes the data covariance. With these definitions, the Fokker-Planck Equation 
(7) is readily seen to be equivalent to the Langevin Equation (3). Multiplying (7) from the left 
by $\eta$ and integrating over phase space  one obtains
\[\frac{d}{dt}<\eta_{n}>=<\Lambda_{n}(t,\eta)>\]
in agreement with the two-component projection of the ensemble average of (3). 

In principle, the advection- and diffusion-terms are determined by the Langevin representation
(3) of the error dynamics. However, particularly for strongly nonlinear dynamics such a
derivation encounters serious formal difficulties and a unique and practical solution to this
problem is currently not known. In view of the typical problem in Earth System Monitoring, 
Belyaev and coworkers \cite{22} propose an alternative, phenomenological determination of these 
parameters. In Earth System Monitoring the time-interval between consecutive samples is
typically short compared to the time scales of the processes under observation. Under these
conditions it becomes possible to consider the model as a black box and determine the
transition (i.e. conditional) probabilities by histogram techniques from model input and output.
To this end, the number $N'$ of all grid points is counted for which $\eta$ has the 
value $y'$ at time $t'$. At the later time $t=t'+\tau$ all former $y'$-points $N$ are counted 
for which $\eta$ now has the value $y$. The conditional probability is then given by the ratio
\[p(t,\eta\,|\,t',\eta')=N/N'.\]
Since $0\leq N\leq N'$, this expression always satisfies the condition
\[0\leq p(t,\eta\,|\,t',\eta') \leq 1\]
necessary for being interpreted as a probability density. With the help of this probability, 
the advection and diffusion parameters are readily obtained by phase space integration according  
to the above formulas.

It is now possible to solve the Fokker-Planck Equation numerically. As a linear parabolic
differential equation in the 2-dimensional unbounded plane, the equation is efficiently 
integrated by the Peaceman-Rachford scheme \cite{25}. For details of this
integration concerning initial conditions, positive definiteness and normalizability of the 
solution see \cite{19}. The resulting probability density determines the error 
covariance at all observational points $X$ according to (6) and the covariance at all model 
grid points is constructed from this expression by interpolation \cite{19}. Using 
this error covariance, the Wiener-Hopf Equation (5) is solved for the Kalman Gain and the model 
is finally updated according to (4).

\vspace{10.mm}
\noindent
{\bf 3.) Ocean State Estimation\\}

The feasibility of operational global ocean state estimation will here be demonstrated by
combining simulations of the numerical circulation model GROB-HOPE with observations of
global sea-surface temperatures (SST) and observed subsurface temperatures from the TAO/TRITON
array for the El Nino year 1997. Besides a globally realistic mean state, the objective of the 
estimate is the improvement of the model's El Nino simulation.

GROB HOPE has 20 layers in the vertical with high resolution of 10 layers in the upper 500m.
In the horizontal, the model uses a spatially inhomogeneous grid obtained from a conformal
transformation of the geographical coordinates. At the present stage of model development 
and availability of computer capacities, the disadvantages of an inhomogeneous horizontal 
grid are easily outweighed by its advantages. For one, polar singularities are avoided by 
transformation of the model poles to a continental site. Secondly, the spatial inhomogeneity 
of the horizontal grid allows high resolution in regions of interest (up to 25 km for the 
Arctic Ocean in the present case) while low resolution is accepted for remote regions (300 km 
near the equator in the present case). This design avoids well-known open boundary problems of 
fine-resolution regional or nested models. While the low-resolution regions provide a 
model-consistent climatology, the high-resolution regions admit even the study of mesoscale 
processes. In spite of this versatility, the machine requirements for GROB HOPE are those of 
a global model with a spatially homogeneous $3^{\circ}\times 3^{\circ}$ grid. This design 
permits a time step of 2.4 hours. With its coarse spatial resolution in the tropics, the
GROB version of HOPE does not especially qualify for El Nino simulation. It is here to be 
shown that assimilation of observations is able to offset these design limitations. Success
in this framework provides a demonstration of the capacities of sequential assimilation.
For operational purposes, on the other hand, data will always be assimilated into the best
model available.

After an initial spin-up period of 2 years with restoring to the 3-dimensional buoyancy 
climatology of WOA the model is integrated from 1948 
to the present with surface forcing derived from the NCEP reanalysis \cite{26}.
Atmospheric data are interpolated onto the GROB HOPE grid and surface buoyancy- and 
momentum-fluxes are calculated by bulk formulae \cite{8} depending on both, the 
atmosphere and the ocean. Hence, the eventual ocean forcing is determined by the particular 
realization of the ocean state by the model while the present ocean-only set-up is unable to
account for a feedback of the ocean on the atmosphere. For a reduction of trends in the deep 
\begin{figure}[!h]
\centerline{\hbox{
\psfig{figure=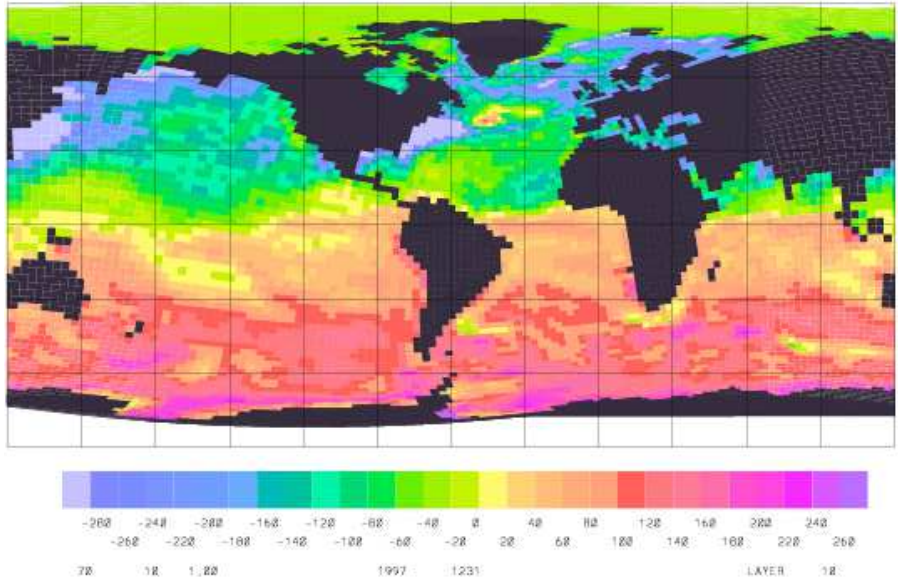,height=12.cm,angle=90,clip=}
}}
\caption{Monthly Mean Surface Heat Flux for December 1997 [W$m^{-2}$]. Control.}
\end{figure}
\begin{figure}[!h]
\centerline{\hbox{
\psfig{figure=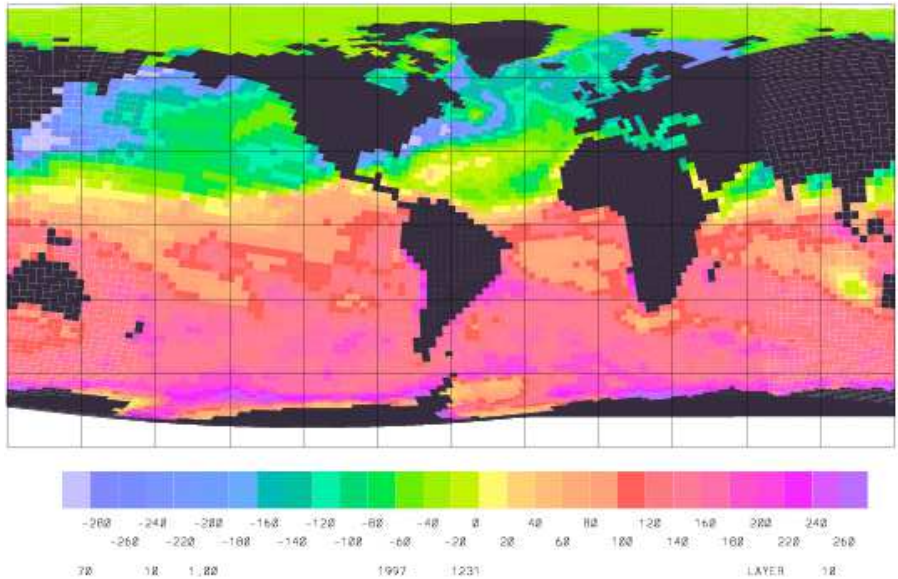,height=12.cm,angle=90,clip=}
}}
\caption{Monthly Mean Surface Heat Flux for December 1997 [W$m^{-2}$]. Nudge.}
\end{figure}
\noindent 
ocean, integration over the NCEP period is repeated. Furthermore, model surface salinities are 
nudged to a mean annual cycle taken from WOA with a time constant of a little over a year 
(385d). Use of a mean annual cycle rather than an annual mean accounts for the seasonal 
variation in the hemispheric distribution of convective activity. With this forcing the model 
is integrated to 31 December 1997. The period from 1 January 1997 to 31 December 1997 is taken 
as the control run in the present experiment and the model state at 31 December 1996 provides 
the initial condition for the assimilation. It is noted that model runs considered here do
not address the prediction problem. Surface data transfer external El Nino information to the
ocean model. 

Fig.2 shows the monthly mean of the net surface heat flux of the control configuration for 
December 1997. This heat flux is determined by atmospheric data from the NCEP reanalysis
and oceanic data from GROB HOPE. The main feature is the characteristic seasonal separation of 
the (southern) summer- and (northern) winter-hemisphere: the ocean gains heat in summer and
loses heat during winter. A particular detail in the North Atlantic is associated with model 
problems in simulating a realistic Gulf Stream path: off the American east coast, the ocean is 
unrealistically warm leading to a pronounced heat loss while the ocean is unrealistically cold in
the region of the so-called North West Corner leading in turn to a pronounced heat gain by the
ocean. Similar aberrations are seen in the Kuroshio region, the confluence of the Malvinas and
Brazil Currents off the South American east coast and for the Agulhas Current near the Cape of 
Good Hope. The paths of these currents are essentially determined by vorticity dynamics and 
mismatches of NCEP derived forcing and model simulation are due to ambiguities in the vorticity 
dynamics of the Primitive Equations. Given the NECP fluxes, GROB HOPE fails to simulate
mesoscale details of the state of the underlying ocean surface realistically.

It will now be shown that nudging of observed SST into the model improves the state estimate
considerably. To this end, GROB HOPE is restarted from 31 December 1996 and daily mean Reynolds
SST of the NCEP data set are inserted into the model's top layer at a time constant of one day.
During this one year integration, model-data incompatibilities do not develop. This is also
true for GROB HOPE runs with SST nudging over the full NCEP period (not shown). Fig.3 depicts
the monthly mean net surface heat flux for December 1997 with SST nudging. In comparison to
fig.2, it is seen that aberrations of the major current systems are significantly reduced and
the estimate of mesoscale features of the state of the sea surface improves without penalty. 

Nudging effects are not confined to the upper ocean alone. In convectively active regions
surface temperature information is rapidly communicated to the abyss. For the present 
integration period of one year the deep ocean remains of course unable to adjust to the
``injected'' information. Nevertheless, with these data and for this model, nudging
becomes a practical option of ocean state estimation by an efficient and yet robust
model-data combination. Other presently available observations are of similar quality:
sea-level data from space-borne altimeters and space-based observations of sea-ice
cover. By nudging observations of this type into a global ocean circulation model, it is 
currently possible to arrive efficiently at a comprehensive and realistic estimate of the 
global state of the sea surface at mesoscale resolution.

For the assessment of the state of the interior ocean consider the equatorial temperature
field during the El Nino episode of 1997/98. Fig.4 shows the temperature difference
Nudge-WOA along the equator for December 1997 where ``Nudge'' refers here to the
GROB HOPE simulation of the global ocean circulation with NCEP forcing and nudging of 
daily SST observations, i.e. the run also portrayed in fig.3. 

In the abyssal Pacific, simulation and observation are seen to differ by typically less
than half a degree. While the simulation is systematically colder than WOA, structural
mismatches do not emerge. The agreement is less satisfactory in the abyssal Indic and
Atlantic. In the near-surface Pacific, the model clearly exhibits the characteristic
El Nino pattern. Relative to the WOA climatology, the eastern and central Pacific
are colder while the West is anomalously warm. Comparison with observed subsurface
temperatures \cite{27} shows that the model simulates the phase of the process quite
realistically. Since phase information is directly provided by forcing data, this model
response is primarily indicative of the consistency of the simulation of near-surface 
wave propagation with with surface boundary conditions.

Other features of fig.4 exhibit a lesser degree of realism. The warm anomaly in the
surface waters of the central Pacific cannot be found in the observational record \cite{27}.
Here, the mixed-layer model of GROB HOPE fails to mix the heat supplied at the surface, 
sufficiently deep into the upper ocean. In the model, heat mainly penetrates to greater
depth by slow diffusion processes. In the ocean, however, these transfers are dominated 
by turbulent mixing. As another consequence of the mixing parameterization, GROB HOPE 
underestimates mixed-layer depths throughout the year and thus fails to account for
Kelvin wave downwelling during El Nino. Thermocline temperatures beneath the
mixed layer are about $2^{\circ}$ Celsius too warm. Here, the model diffuses too much
heat to depths of approximately 500m in the eastern equatorial Pacific which penetrates
westward at approximately 250m. This mismatch is the result of unrealistically strong
downward diffusion of heat and unrealistically weak upwelling of cold waters.

For Primitive Equation models, vertical transfers pose a greater problem. Nonhydrostatic
mixing processes have to be parameterized and such parameterizations are by no means trivial.
The mixed-layer model of GROB HOPE is tuned to yield realistic mixing depths at moderate
latitudes and compromises for the equatorial mixed layer are accepted. The alternative 
would be a far more complex and machine-intensive mixed-layer model. Moreover, vertical
velocities are determined from mass conservation, independent of the momentum budget.
Possible problems and ambiguities are smeared out by diffusion. Hence, models have a
tendency to use diffusion where space-time- and phase-space-characteristics of the
real ocean are determined by advection and propagation.

Sequential assimilation of subsurface temperatures improves this state estimate 
significantly. Subsurface temperature data are taken from the TAO/TRITON
array which consists of approximately 70 moorings in the tropical Pacific between 
$8^{\circ}S$ and $8^{\circ}N$. The buoys record a number of atmospheric parameters,
sea surface temperatures and subsurface temperatures at 10 irregularly spaced depths in 
the upper 500m. Records are transmitted to shore in real-time via the ARGOS satellite 
system. TAO/TRITON has become one of the most successful ground-based ocean observatories 
for two major reasons. In the first place, the relatively quiescent tropical waters allow 
the long-term deployment of buoys. For the more energetic high latitude oceans, the physical
lifetime of a similar array would be significantly shorter than time scales of ocean
processes of interest. Secondly, the observed variability is readily interpreted in terms
of Matsuno's theory. Since 1985, the TAO/TRITON array has become an integral part of 
operational services as well as ocean and climate research.

In contrast to the surface boundary condition, nudging of regional TAO/TRITON data is not 
an option. This array is physically embedded in the global ocean circulation and its
data interact physically with their neighborhood in space-time as well as in phase space. 
In the Fokker-Planck picture of the Kalman Filter this is accounted for by phase space 
advection and diffusion which are updated at every assimilation step. To this end, model 
integration is halted at the end of each month and the model temperature field is updated 
by observed TAO temperatures. After the update, model integration resumes and continues
for one month when model temperatures are updated again. Thus, model operation is 
constrained to the vicinity of the observed state of the ocean.

\begin{figure}[!h]
\centerline{\hbox{
\psfig{figure=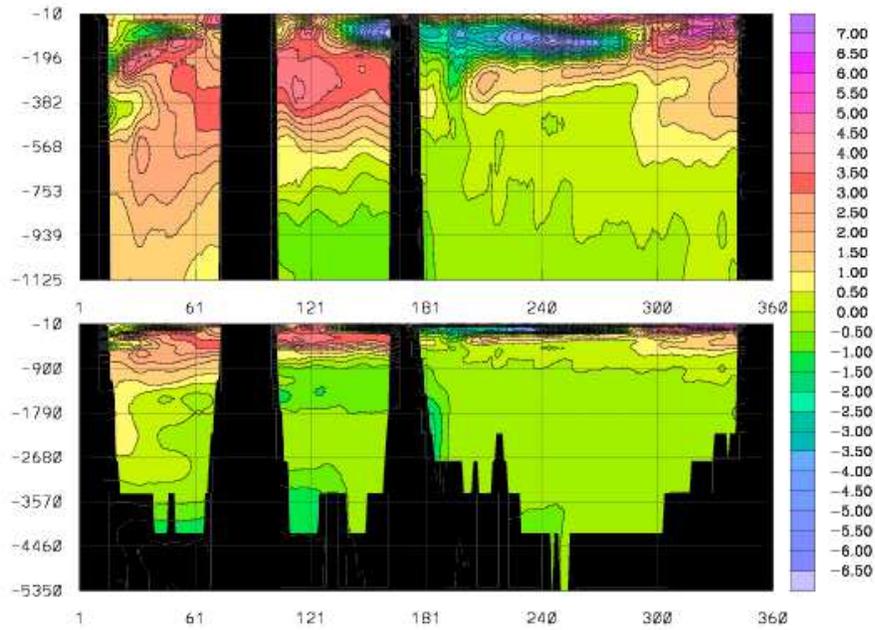,height=12.cm,angle=90,clip=}
}}
\caption{Monthly Mean Temperatures on the Equator. December 1997. Nudge-WOA.}
\end{figure}

\begin{figure}[!h]
\centerline{\hbox{
\psfig{figure=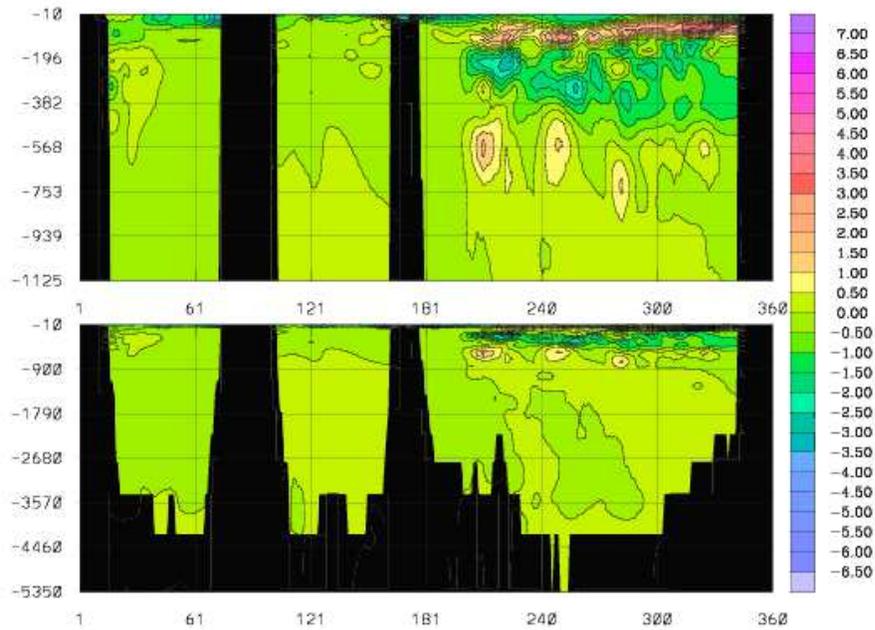,height=12.cm,angle=90,clip=}
}}
\caption{Monthly Mean Temperatures on the Equator. December 1997. Assimilation-Nudge.}
\end{figure}

Fig.5 shows the monthly mean temperature difference Assimilation-Nudge along the equator 
for December 1997. The data are seen to have three major effects on the estimate: the
surface becomes colder, the mixed layer warmer and the thermocline colder. These 
modifications lead to a significantly higher degree of realism for the estimate.
Assimilation ensures that heat supplied at the surface, is uniformly mixed into the 
upper layer and cold water is upwelled into the thermocline. In response to the data
information, the model replaces diffusion dominated dynamics with mixing and advection.
However, it is also seen that the assimilation still exhibits some pockets of warm water 
below 500m although far less than fig.4. Primarily, these pockets are a consequence of 
the lower boundary condition chosen for vertical transition probabilities: the model
is assumed to be true at 500m. Obviously, there is room for improvement.
 
\begin{figure}[!h]
\centerline{\hbox{
\psfig{figure=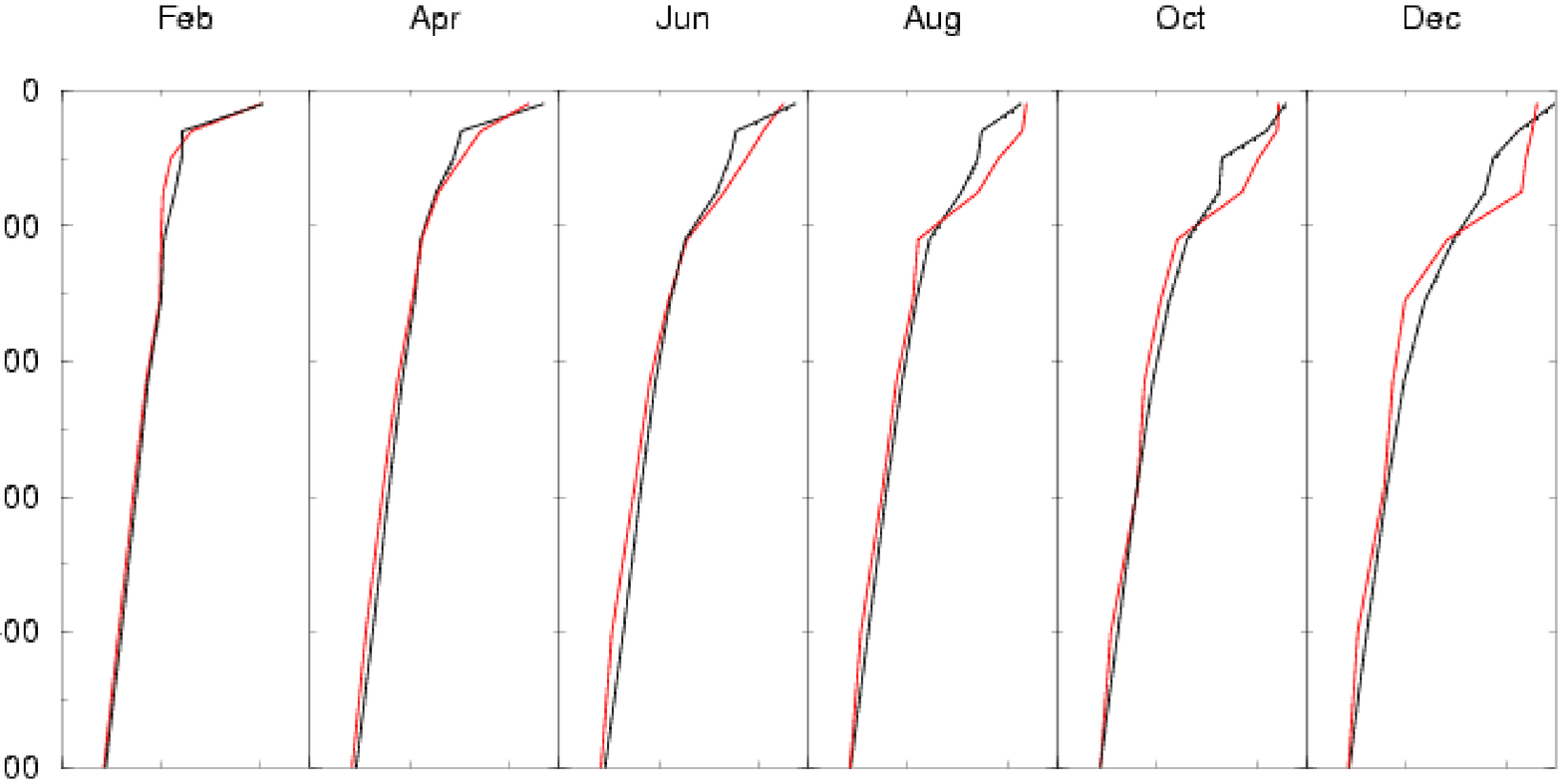,height=12.cm,angle=270,clip=}
}}
\caption{Monthly Mean Temperature Profiles at $260^{\circ}$, $0^{\circ}$ in 1997. 
Nudge (black), Assimilation (red).}
\end{figure}

A different view of these data effects is given in fig.6. The figure shows a time series of
monthly mean temperature profiles at a location in the eastern equatorial Pacific for 1997.
Simulated (black) and assimilated (red) profiles are compared. The simulation is clearly
diffusion dominated, unable to produce a mixed layer and leaking too much heat into the
thermocline. The assimilation also fails to produce well-defined mixed layers during the
first part of 1997. Before the arrival of the downwelling Kelvin wave in the eastern
Pacific, mixed layers here are shallow (typically 25m). Their absence in the assimilation
during the first part of the year is a consequence of the poor vertical resolution of
GROB HOPE. With the arrival of the Kelvin wave, assimilation produces the characteristic
signature of turbulent mixing in the upper ocean with realistic mixed-layer depth. At the
same time, the thermocline is cooled by upwelling of colder water and temperature
gradients at the mixed-layer base increase.

It is mentioned (corresponding plots not shown) that the model extrapolates data information
beyond the temperature field in the TAO/TRITON region. By December 1997, temperatures as
far as $30^{\circ}$ latitude on both hemispheres are modified by the assimilation. In
phase space, model dynamics extrapolate temperature observations onto sealevel, salinity
and the velocity field and improve the estimate for the complete state of the ocean. On
the other hand, without continued assimilation the model loses memory of the data
information from the upper equatorial Pacific after a time period of three to six months.

\vspace{10.mm}
\noindent
{\bf 4.) Summary\\}

Societal needs of climate monitoring call for the installation of global ocean state
estimation in the framework of an operational service similar to national weather
prediction agencies. Novel ocean observation techniques provide a global data base
for such estimates. For a number of parameters, these observations are available almost 
in real-time and at mesoscale resolution. Assimilators utilize numerical circulation
models to dynamically extrapolate measurements in space-time and phase space and, at
the same time, constrain model uncertainties.

Given the data volume in ocean observation and modeling, requirements in computing
resources are high. For efficiency, it is significant that model and data quality
is sufficiently developed to allow nudging the model to essentially unprocessed
data at short time constants. This applies particularly to the sea surface where
data wealth is largest and determines primarily the surface forcing with a high
degree of realism. The forcing problem that plagued ocean model development 20 
years ago has been resolved.

Dynamical extrapolation of interior ocean data requires advanced assimilation
techniques. Monitoring problems are best addressed by sequential methods such as
the Kalman Filter. Typically, observations do not coincide with the corresponding
model solution and may even be incompatible with model dynamics. The Kalman Filter
finds the initial value for model restart that is both, as close as possible to the 
data and compatible with model dynamics. To this end, the phase space representation
of the filter solves (a large number of) simple, 2+1 dimensional, linear
Fokker-Planck equations which represent model dynamics in terms of phase space
advection and diffusion. Determination of these parameters by an elementary
histogram method circumvents a number of highly complex, but essentially technical
issues of the stochastics of nonlinear systems. For numerical analysis, the 
method proves efficient and reliable.

At this time, models alone are generally unable to simulate the global ocean 
circulation with satisfactory realism. However, they do capture large-scale
features such as gyres, water masses and their seasonal and longer-term 
variation quite realistically if they are determined by large-scale features
of topography and external forcing. Problems typically emerge with the dynamical
control of the density field. In the present study, this is demonstrated for
equatorial mixing and upwelling and the paths of major ocean currents. On the 
other hand, the assimilation has shown that models are not antagonistic to 
(temporary) operation in closer vicinity of the data. At this development stage, 
model operation in the assimilation mode is capable of delivering practically 
relevant global ocean state estimates provided a continuous inflow of data is 
guaranteed.

Some improvement in model performance is readily obtained by fairly simple measures.
Higher vertical resolution is oftentimes possible within the framework of given
computing resources. Significant increase of the overall horizontal resolution
requires generally an upgrade of computer resources. For the performance of HOPE at
higher spatio-temporal resolution, see \cite{8}. Moreover, the modeling community
continuously develops increasingly appropriate parameterizations of subscale 
processes and models are updated accordingly.

On the other hand, as long as models are unable to account for basic laws of nature 
such as the first law of motion, long-term projections with little or no data input
after initialization will remain questionable. In this context, it is noted that the 
``Newtonization'' of Richardson's equations is actually more or less trivial: 
vertical integration of the Primitive Equations for an incompressible
(multilayer) fluid leads again to a Newtonian set of equations of motion. In this
framework, vertical variability appears as internal variability of the spatially strictly
2-dimensional fluid. Minor questions arise from the hierarchy problem that results from
vertical integration of the nonlinear advection term. Since stratification is represented
by the multilayer structure, low-order cut-offs of the advection hierarchy suffice for
most purposes of circulation modeling. This shallow water approach to the circulation
problem can be formulated with geometric-dynamic integrity and energetic consistency in
both, the viscous nonrotating limit and the ideal rotating limit. The theory also
admits comprehensive analytical studies of wave-circulation interaction. Currently, 
physically consistent global circulation models on the basis of shallow water theory 
are not available.

\end{document}